\documentclass[10pt,aps,prd,nofootinbib,notitlepage,superscriptaddress,twocolumn]{revtex4-1}

\usepackage{amsfonts}
\usepackage{amsmath}
\usepackage{xcolor}
 \usepackage{cancel}
 
\newcommand{\ee}{\mathrm{e}}
\newcommand{\dd}{\mathrm{d}}
\newcommand{\Q}{\mathbb{Q}}

\usepackage[colorlinks=true,hyperfootnotes=true,citecolor=cyan]{hyperref}

\begin{document}

\title{Physical nonviability of $f(\Q)$ in the scalar-tensor representation}

\author{Jose Beltr{\'a}n Jim{\'e}nez}
\email{jose.beltran@usal.es}
\affiliation{Departamento de Física Fundamental and IUFFyM, Universidad de Salamanca, E-37008 Salamanca, Spain}

\author{Alejandro Jim\'enez Cano}
\email{alejandro.jimenez.cano@upm.es}
\affiliation{Escuela T\'ecnica Superior de Ingenier\'ia de Montes, Forestal y del Medio Natural, Universidad Politécnica de Madrid, 28040 Madrid, Spain}

\author{Tomi S. Koivisto}
\email{tomi.koivisto@ut.ee}
\affiliation{Laboratory of Theoretical Physics, Institute of Physics, University of Tartu, W. Ostwaldi 1, 50411 Tartu, Estonia}
\affiliation{National Institute of Chemical Physics and Biophysics, R\"avala pst. 10, 10143 Tallinn, Estonia}

\begin{abstract}
We show the known pathological character of $f(\Q)$ gravity in its scalar-tensor representation.
\end{abstract}

\maketitle

\section{Introduction}
\label{sec:intro}

Teleparallel formulations of General Relativity (GR) have been widely explored, as they allow gravity to be described in terms of torsion and/or nonmetricity rather than curvature, while remaining dynamically equivalent \cite{Trinity, BeltranJimenez:2019odq}. In particular, \emph{symmetric teleparallel gravity} \cite{Nester:1998mp,Adak:2005cd} provides a geometrical framework in which gravity is encoded in the nonmetricity tensor associated to the teleparallel connection $\nabla$ and the spacetime metric $g_{\mu\nu}$:
\begin{equation}
    Q_{\mu\nu\rho} \equiv \nabla_\mu g_{\nu\rho}\,.
\end{equation}
In this framework, GR can be recovered through the nonmetricity scalar
\begin{align} \label{Qscalar}
    \Q &\equiv \frac{1}{4} Q_{\mu\nu\rho} Q^{\mu\nu\rho} - \frac{1}{2} Q_{\mu\nu\rho} Q^{\nu\mu\rho} \nonumber\\
    &\qquad -\frac{1}{4} Q_{\mu\nu}{}^\nu Q^{\mu\rho}{}_\rho + \frac{1}{2} Q_{\mu\nu}{}^\nu Q_\rho{}^{\rho\mu}\,,
\end{align}
which plays a role analogous to that of the Ricci scalar in GR (their difference is in fact a total derivative) and constitutes the building block of a large class of modified gravity models, including $f(\Q)$ theories and their generalizations \cite{BeltranJimenez:2017tkd, BeltranJimenez:2018vdo, BeltranJimenez:2019tme}. 

It was argued in \cite{Gomes2023} that the general framework of symmetric teleparallel theories is prone to the presence of ghost-like instabilities due to the breaking of the gauge symmetries that render the graviton self-interactions stable. 

An exception to this general result has been put forward in \cite{Bello-Morales:2024vqk} where it was shown that a partial breaking of diffeomorphism invariance down to its volume-preserving subgroup leads to ghost-free theories. In fact, they were shown to be equivalent to scalar-tensor theories with a constant parameter that arises as an integration constant, very much like in unimodular gravity. However, barring these exceptional situations, the generic argument remains valid and it certainly applies to $f(\Q)$ theories, as it was explicitly shown in \cite{Gomes2023}, and other extensions that we will discuss here.

The pathological character of $f(\Q)$ gravity has been known since its inception \cite{BeltranJimenez:2019tme},  where maximally symmetric backgrounds were already shown to be strongly coupled. This compromises the physical relevance of the whole family of cosmological solutions in the small-scale limit, as well as that of spherically symmetric configurations that are asymptotically flat or (A)dS. It remained open, however, whether these pathologies extend to more general cases. In \cite{Gomes2023}, it was explicitly shown that all spatially flat cosmological solutions are either strongly coupled or contain a ghost-like instability.\footnote{
    The sudden singularities found in Ref.~\cite{Guzman:2024cwa} may be viewed as a background-level symptom of the same instability.}
It is important to note that the strongly coupled branches of solutions are not ghost-free, but the ghost is expected to be one of the strongly coupled degrees of freedom.

A crucial point missed in much of the recent literature is that Ref.~\cite{Gomes2023} already fixes the local spectrum of $f(\Q)$ gravity: it propagates seven degrees of freedom. Strong coupling does not remove the extra modes, but merely hides some of them at linear order, including a ghost. Phenomenological applications of generic $f(\Q)$ models are therefore physically irrelevant (unless the ghost is removed, and no mechanism is currently known to achieve this).

Here we perform the scalar-tensor analysis anticipated in  Ref.~\cite{Gomes2023}, and thereby close a potential loophole. As this brings the stability issues once again to the fore, we take the opportunity to clarify the pathological degrees of freedom, assess their robustness under field redefinitions and matter-sector extensions, and address some misconceptions found in the literature.

\section{Cosmological perturbations of \texorpdfstring{$f(\Q)$}{f(Q)} in scalar-tensor form}
\label{sec:EinframeScalar}

We consider $f(\Q)$ theories coupled to a scalar field as the matter sector so the system is described by the following action:
\begin{equation}
    \mathcal{S} = - \frac12 \int\dd^4x \sqrt{-g} \ \left[ f(\Q) + (\partial \chi)^2 +2V(\chi)\right] \,.
    \label{eq:fQ}
\end{equation}
The basic variables are the matter field $\chi$, the metric $g_{\mu\nu}$ and the fields $\xi^\mu$, which define the symmetric teleparallel connection through
\begin{equation}
    \Gamma^\alpha{}_{\mu \beta} = \frac{\partial x^\alpha}{\partial \xi^\rho} \frac{\partial^2 \xi^\rho}{\partial x^\mu \partial x^\beta}\,.
\end{equation}
Following the steps provided in \cite{BeltranJimenez2021},\footnote{
    Notice that our $\varphi$ corresponds to the $\phi$ in that paper.
    }
it is possible to introduce an auxiliary scalar field $\varphi$ such that, $\ee^{2\varphi} = \dd f /\dd \Q$ on shell. After redefining the metric as $g_{\mu\nu} = \ee^{-2\varphi} q_{\mu\nu}$,  the action \eqref{eq:fQ} can be brought to the scalar-tensor form in terms of the new set of variables $\{q_{\mu\nu},\xi^\mu,\varphi,\chi\}$:
\begin{align}
    \mathcal{S}_{\text{ST}} =\frac12 \int\dd^4x\,\sqrt{-q}\ &\Big[R(q)+6q^{\mu\nu}\partial_\mu \varphi\partial_\nu \varphi\nonumber\\
    & -2 \big(q^{\alpha\beta}q^{\mu\nu}-q^{\alpha\mu} q^{\beta\nu}\big)\partial_\alpha\varphi\nabla_\beta q_{\mu\nu}  \nonumber \\
    & - \ee^{-2\varphi} q^{\mu\nu}\partial_\mu \chi \partial_\nu \chi +\mathcal{U}(\varphi, \chi)\Big]
    \label{eq:fQ_ST}
\end{align}
for a certain function $\mathcal{U}$. Here $R(q)$ is the Ricci scalar associated to the new metric $q_{\mu\nu}$. 

Note that the symmetric teleparallel connection enters the scalar-tensor action \eqref{eq:fQ_ST} only linearly through $\nabla_\beta q_{\mu\nu}$,  and, thus, so do second-order derivatives of $\xi^\mu$. This means that second order derivatives of the connection fields $\xi^\mu$ mix with derivatives of the conformal scalar field $\varphi$ and, therefore, contribute to the kinetic sector. This already provides a first indication of potential pathologies. Of course, a more careful analysis requires first integrating out all auxiliary fields, since such integrations may alter the signs in the kinetic matrix. This procedure was carried out in \cite{Gomes2023} using the formulation \eqref{eq:fQ} expanded around cosmological backgrounds. The results show that the presence of ghosts is unavoidable. Note that we are using cosmological backgrounds as proxies to make manifest a more severe problem that is intrinsic to the theory.

It is amusing to notice that we can express the derivative mixing in \eqref{eq:fQ_ST} in terms of Levi-Civita tensors as 
\begin{equation}
    2\Big(q^{\alpha\beta}q^{\mu\nu}-q^{\alpha\mu}q^{\beta\nu}\Big)\partial_\alpha\varphi\partial_\beta q_{\mu\nu} = -\epsilon^{\alpha\mu\lambda\kappa}\epsilon^{\beta\nu}{}_{\lambda\kappa}\partial_\alpha\varphi\partial_\beta q_{\mu\nu},
    \label{eq:epsilonepsilon}
\end{equation}
which resembles the structure of e.g. Lovelock invariants or massive gravity theories. This might give some hope that the interactions have some special structure preventing the propagation of a ghost. In fact, the main pathology of $f(\Q)$ theories could be interpreted as a severe version of a Boulware-Deser ghost \cite{Boulware:1972yco} so one may be hopeful that $f(\Q)$ might avoid the presence of ghosts in a similar manner to massive gravity given the similarity of the interactions written in terms of Levi-Civita tensors. We will come back to the analogy with massive gravity in Sec. \ref{sec:conclusions}, but the main lesson extracted from the explicit computation in \cite{Gomes2023} is that $f(\Q)$ in turn suffer from the presence of the ghost.

Here, we perform the analogous analysis in the scalar-tensor theory defined by \eqref{eq:fQ_ST}. We therefore consider the cosmological background (including the matter field)
\begin{align}
    q_{\mu\nu}(t,x^i) &= \mathrm{diag}\big(-n(t)^2,\, a(t)^2,\, a(t)^2,\, a(t)^2\big)\,,\label{eq:metric}\\
    \xi^\mu(t,x^i)&= \left( \xi(t) - \frac{\alpha_\text{I}}{2} \sigma_0 \lambda |\vec{x}|^2 , \ (\alpha_\text{II}\lambda \xi(t) + \sigma_0) x^i \right)\,,\\
    \varphi(t,x^i)&= \bar{\varphi}(t)\,,\\
    \chi(t,x^i)&= \bar{\chi}(t)\,,
\end{align}
which is fully characterized by the functions $\xi(t)$, $n(t)$, $a(t)$, $\bar{\varphi}(t)$ and $\bar{\chi}(t)$, and the parameters $\sigma_0$ and $\lambda$. The three branches of $\xi^\mu$ compatible with the cosmological symmetries of \eqref{eq:metric} in symmetric teleparallel gravity correspond to each of the following possibilities: $(\alpha_\text{I}, \alpha_\text{II})=(0,0)$, $(1,0)$ or $(0,1)$, which are called trivial branch, branch I and branch II, respectively.\footnote{
    See \cite{Gomes2023cosmo} for more details on how these three branches are constructed. Appendix A of that paper discusses the relation to the alternative parameterisation of the connection coefficients given in \cite{Hohmann:2020zre}.
}

We study inhomogeneous perturbations in terms of $\alpha_{\rm I}$ and $\alpha_{\rm II}$, and with the background metric expressed in conformal time $\eta$ (i.e., $n=a$). We choose a gauge with unperturbed $\xi^\mu$ for which $\delta\Gamma^\alpha{}_{\mu\nu}=0$ and express the perturbations of the rest of the field variables as follows:
\begin{align}
    \delta q_{00} &= - 2 a^2 \phi  \,,\\
    \delta q_{0i} &= a^2 (\partial_i B + B_i)\,,\\
    \delta q_{ij} &=  a^2\left[ -2 \psi \delta_{ij}+ \partial_i\partial_j E + \frac12(\partial_{i}E_{j}+\partial_{j}E_{i})+ h_{ij}\right] \,,\\
    \delta\varphi &= \zeta\,,\\
    \delta\chi &= \pi\,,
\end{align}
with $\delta^{ij}h_{ij}=0$ and $\partial_iB^i=\partial_iE^i=0=\partial_i h^{ij}$. We follow the steps in \cite{Gomes2023} (as well as the conventions used there) to obtain the following expressions for the tensor and vector sectors in Fourier space:
\begin{align}
    \mathcal{S}^{(2)}_{\rm ten} &=\frac12\sum_{\alpha=+,\times}\int \dd \eta \dd^3k\ a^2 \left[\vert h'_\alpha\vert^2-\left(k^2 + \alpha_{\rm I} \frac{\xi' J}{a^2} \right)\vert h_\alpha\vert^2\right] \,,\\
    \mathcal{S}^{(2)}_{\rm vec} & =\frac12 \int\dd \eta\dd^3k\ \alpha_{\rm I}\xi' J\left[ \frac{\vert\vec{E}'\vert^2}{1+ \dfrac{\xi' J}{a^2 k^2}} - k^2 |\vec{E}|^2\right]\,,
\end{align}
where $\vec{B}$ has already been integrated out from the second one as it enters as an auxiliary field. Here, we are calling
\begin{equation}
    J\equiv  -\frac{4\lambda \sigma_0 a^2 \varphi'}{\xi'^2}\,,
\end{equation}
which is a conserved quantity for the branch I.

The expressions, and hence the conclusions, are identical to those obtained in Ref.~\cite{Gomes2023} in the original $f(\Q)$ frame, after the
replacements $f_{\Q}\to 1$ and $J_{\rm I}\to J$.
To sum up, we observe a potential tachyonic instability in the tensor sector for some branch I backgrounds, ghost degrees of freedom in the vector sector in the high-momentum regime unless $J \xi' >0$, and strong coupling of the vector degrees of freedom around the trivial branch and branch II (since the whole Lagrangian is proportional to $\alpha_\text{I}$). 

Finally, we turn our attention to the scalar sector. The perturbations $\phi$ and $B$ are non-dynamical. The former can be integrated out, whereas this is only possible for the latter around branch I, since $B$ enters its equation of motion multiplied by $\alpha_{\rm I}$ (in other words, $B$ is a Lagrange multiplier in the trivial branch and in branch II). If we assume the background corresponds to the branch I configuration, upon integrating out these fields, we can write the scalar sector as:
\begin{equation}
    \mathcal{S}{^{(2)}_{\rm scal}}=\frac12\int\dd \eta\dd^3k\ a^2 \left[ \Phi'\mathcal{K}_{\rm s}\Phi'{}^\dagger+\Phi'\mathcal{N}_{\rm s}\Phi^\dagger-\Phi\mathcal{V}_{\rm s}\Phi{}^\dagger+\text{c.c.}\right]
\end{equation}
with $\Phi = (\zeta, \psi, E, \pi)$ and c.c. stands for complex conjugate. The determinant of the kinetic matrix has the following asymptotic behavior:
\begin{equation}
    \det{\mathcal{K}_{\rm s}} \simeq - 2 \frac{J}{\xi'} \left(\frac{6\lambda^2 \sigma_0^2 a \, \ee^{-\bar{\varphi}}}{8\lambda\sigma_0 aa' + J \xi'^2}\right)^2 k^2
    \qquad (k^2 \to \infty) \,.
\end{equation}
Provided the vector sector is ghost-free ($J$ and $\xi'$ have the same sign), this quantity is always negative, thus indicating the presence of a ghostly scalar degree of freedom in the high-momentum regime. Hence, the scalar-tensor reformulation reproduces the original obstruction \cite{Gomes2023} in a different set of variables.

\section{Propagating degrees of freedom}

Let us pause to emphasize that the linear spectrum contains 2 tensor + 2 vector + 4 scalar dofs. Since we are considering a matter sector corresponding to a scalar field, we conclude that the scalar-tensor representation of $f(\Q)$ theories propagates 7 dofs, again in agreement with the findings in \cite{Gomes2023}. It is important to notice that all these seven dofs are propagating at linear order since we have integrated out all non-dynamical fields so the resulting quadratic action only contains dynamical dofs. This means that these seven dofs also constitute the dynamical dofs at full non-linear order.

There might remain some doubt as to whether we have unveiled all the propagating dofs or additional ones might appear around less symmetric backgrounds. It is however easy to see from the scalar-tensor representation \eqref{eq:fQ_ST} that 7 is in turn the maximum number of dofs we could have obtained. Without loss of generality, we can work in the coincident gauge where the connection trivialises and, hence, all derivatives become partial derivatives. In this gauge, the first and third lines in \eqref{eq:fQ_ST} simply correspond to standard GR coupled to two scalar fields and the counting of dofs is the usual. Thus, we only need to analyse the second line in \eqref{eq:fQ_ST} because those terms are the Diff-breaking terms responsible for the additional dofs. We will employ its expression \eqref{eq:epsilonepsilon} and will focus on time derivatives. We then obtain
\begin{align}
   2 \mathcal{S}_{\cancel{\mathrm{Diff}}}&=\int\dd^4x\sqrt{-q}\,\epsilon^{\alpha\mu\lambda\kappa}\epsilon^{\beta\nu}{}_{\lambda\kappa}\partial_\alpha\varphi\partial_\beta q_{\mu\nu}\nonumber\\
    &=\; \int\dd^4x\sqrt{-q}\,\epsilon^{0imn}\epsilon^{0j}{}_{mn}\partial_0\varphi\partial_0 q_{ij}\nonumber\\
    &\quad +\int\dd^4x\sqrt{-q}\,\epsilon^{0imn}\epsilon^{j0}{}_{mn}\partial_0\varphi\partial_j q_{i0}\nonumber\\
    &\quad + \int\dd^4x\sqrt{-q}\,\epsilon^{i0mn}\epsilon^{0j}{}_{mn}\partial_i\varphi\partial_0 q_{0j}\nonumber\\
    &\quad +\text{terms without time derivatives}.
\end{align}
This expression clearly shows that $q_{00}$ does not contain any time derivatives and the time derivatives of $q_{0i}$ only enter linearly and without mixing with other time derivatives. Since the other sector, which describes GR coupled to two scalar fields, does not contain time derivatives of $q_{00}$ and $q_{0i}$ because these can be written in terms of the lapse and shift ADM variables that ensure the Hamiltonian and momentum constraints, those two components of the metric are auxiliary fields that can be integrated out. Observe that the auxiliary variables $q_{00}$ and $q_{0i}$ are directly related to $\phi$ and the pair $B$ and $B_i$, respectively, which are precisely the ones that have been integrated out in the previous section. It is important to notice that, after integrating out these non-dynamical fields, no higher derivative terms for other fields are generated. All in all, this means that, out of the 10 components of $q_{\mu\nu}$ only its spatial components $q_{ij}$ can correspond to dynamical fields. These 6 spatial components together with the conformal mode $\varphi$ then make up a total of 7 dofs which, then, is the maximum number of dofs the theory can propagate. This is of course not an exhaustive analysis of the propagating dofs in the theory, but the fact that we have found a background that saturates this maximum number of dofs means that that is in fact the number of dofs of the theory.

\section{Generalizations}
\label{sec:generalizations}

Now, as promised in \cite{Gomes2023}, we present a more detailed analysis of the following case, which generalizes the previous result:
\begin{equation}
    \mathcal{S} = \int\dd^4x \sqrt{-g} \left[-\frac12 f(\Q,\chi)-\frac12(\partial\chi)^2-V(\chi)\right]\,.
\end{equation}
We follow closely the procedure devised in \cite{BeltranJimenez2021}. We first introduce two scalar auxiliary fields $Y$ and $\varphi$ to construct the equivalent action:\footnote{
    The equations of motion of $Y$ and $\varphi$ can be recast as:
    \[
    Y= \Q \,,\qquad \ee^{2\varphi} = \frac{\partial f(\Q, \chi)}{\partial \Q}\,,
    \]
    which indeed lead to the original action.
    }
\begin{align}
    \mathcal{S} = -\frac12\int\dd^4x \sqrt{-g} &\Big[ f(Y,\chi) + (\Q - Y)\ee^{2\varphi} \nonumber \\
    & \qquad + (\partial\chi)^2 + 2 V(\chi)\Big]\,.
\end{align}
Now we assume that there is a solution of the equation of motion of $Y$, say $Y=Y_0(\chi, \varphi)$. We plug this back into the action and get
\begin{align}
    \mathcal{S} &= \frac12\int\dd^4x \sqrt{-g} \left[ -\ee^{2\varphi}\Q  - (\partial\chi)^2 + \tilde{\mathcal{U}}(\varphi, \chi)\right]
\end{align}
with
\begin{equation}\label{eq:tildeU}
    \tilde{\mathcal{U}}(\varphi, \chi) \equiv Y_0(\chi, \varphi)\ee^{2\varphi} - f(Y_0(\chi, \varphi),\chi) - 2 V(\chi)\,.
\end{equation}
Finally, we express $\Q$ in terms of the Ricci scalar $R(g)$, integrate by parts the extra term so as to transfer the derivatives onto the factor $\ee^{2\varphi}$, and perform the conformal transformation $g_{\mu\nu} = \ee^{-2\varphi} q_{\mu\nu}$. The resulting action takes the scalar-tensor form:
\begin{align}
    \mathcal{S}_{\text{ST}} = \frac{1}{2} \int \dd^4x \,\sqrt{-q}\ & \Big[R(q) + 6 q^{\mu\nu} \partial_\mu \varphi \partial_\nu \varphi\nonumber\\
    & -2 \big(q^{\alpha\beta} q^{\mu\nu} - q^{\alpha\mu} q^{\beta\nu}\big) \partial_\alpha\varphi \nabla_\beta q_{\mu\nu}  \nonumber \\
    & - \ee^{-2\varphi} q^{\mu\nu} \partial_\mu \chi \partial_\nu \chi +\tilde {\mathcal{U}}(\varphi, \chi)\Big]\,,
    \label{eq:fQchi_ST}
\end{align}
which is identical to \eqref{eq:fQ_ST}, but with $\tilde{\mathcal{U}}$ playing the role of the potential $\mathcal{U}$. Hence, the kinetic sector remains unaltered, and the analysis in Sec.~\ref{sec:EinframeScalar} applies equally to this case.

Having shown that the previous conclusions are robust under this reformulation, we now ask whether they persist for more general classes of theories. To this end, we consider a theory of the form:
\begin{equation}
    \mathcal{S} = - \frac{1}{2} \int \dd^4x \sqrt{-g} \ f(\Q,\chi,(\partial \chi)^2) \,.
\end{equation}
The exact same procedure leads again to the form \eqref{eq:fQ_ST} although now  $\tilde{\mathcal{U}}$ depends also on $(\partial \chi)^2$, therefore the kinetic part of the scalar sector changes in general. The tensor and vector sectors, however, are unaffected by this modification. Thus, the vector mode $\vec{E}$ is ghostly if $J \xi' \leq 0$. Although the strong coupling can be avoided by moving to another branch, the ghost is expected to persist in the scalar sector, since the higher-order derivative term remains present for generic $f$.

These conclusions generalize straightforwardly to multi-scalar extensions and even to cases in which the derivatives do not enter through the combination $(\partial \chi)^2$. The situation can become even more problematic if higher-order derivatives are allowed. Thus, to make the point clear, except for exceptional cases in which the fields possess a special (usually accidental) symmetry, \emph{a generic theory} of the type $f(\Q,\, \chi^A,\, \partial_\mu \chi^A, \ldots,\, \partial_\mu \cdots \partial_\nu \chi^A )$, where $\{\chi^A\}$ is any set of scalars, exhibits the problems we have just pointed out in the vector and scalar degrees of freedom. This includes as a particular case the theory $f(\Q,T)$ (where $T$ stands for the trace of the energy-momentum tensor) or the theories $f(\Q,\mathcal{B})$ with $\mathcal{B}=R(g)-\Q$.\footnote{
    Obviously, if we choose $f(\Q,\mathcal{B})=f(\Q+\mathcal{B})$, we recover the standard $f(R)$ theories that propagate one additional healthy scalar field.
    }

The same conclusions extend as well to theories featuring more general geometric invariants than $\Q$. 
The generic quadratic, parity-even and second-order derivative symmetric teleparallel gravity action consists of the 5 independent scalars constructed from the quadratic contractions of the nonmetricity tensor. The quadratic 5-parameter theory suffers from pathologies \cite{BeltranJimenez:2018vdo}, see also \cite{Ganz:2025ydt}, unless restricted either to the particular linear combination $\Q$ given at (\ref{Qscalar}), or to the transverse-diffeomorphism-preserving special case \cite{Bello-Morales:2024vqk}. Nonlinear deformations of the latter are obstructed by the same mechanism responsible for the pathology of the $f(\Q)$ model. 

Recent analyses of theories that combine extra scalar fields with the general five-parameter nonmetricity sector likewise support this overall picture \cite{Chen:2025mhu}.

\section{Discussion}
\label{sec:conclusions}

In this note we have shown that the scalar-tensor representation of $f(\Q)$ gravity does not evade the known strong-coupling/ghost obstruction, and that the same problem persists in natural extensions. We now close by discussing several further putative loopholes and explain why none removes the instability.

~

\textbf{The role of the gauge}.
The connection in $f(\Q)$ theories, as well as in extensions of the types discussed in Sec.~\ref{sec:generalizations}, can be trivialised by going to the\footnote{
    The coincident gauge $\Gamma^\alpha{}_{\mu\beta} = 0$ is not unique, in the sense that the corresponding $\xi^\mu$ is determined only up to a global affine transformation ($\xi^\mu \to A^\mu{}_\nu \xi^\nu + b^\mu$ with constant $A^\mu{}_\nu$ and $b^\mu$).
}
coincident gauge and this does not alter any of the theory dynamics. The fundamental reason is that the symmetric teleparallel connection just introduces St\"uckelberg fields ($\xi^\mu$) associated to diffeomorphisms and the coincident gauge simply amounts to identifying those fields with the coordinates ($\xi^\mu = x^\mu$), something that we can always do. The only subtle point here is related to topological properties, but this does not affect e.g. the local degrees of freedom. As a matter of fact, all this is true for any theory formulated in the symmetric teleparallel framework, not just $f(\Q)$. This was already discussed in some detail and clarified in e.g. \cite{BeltranJimenez:2022azb,Jensko:2024bee}, but some confusion seems to persist in the literature. We stress once more that
\emph{there is no difference between the theory expressed in the coincident gauge and its covariant formulation}. 

~

\textbf{Background vs intrinsic ghost}. 
That there is a severe pathology in generic $f(\Q)$ theories is already evident from the structure of the full (unperturbed) action \eqref{eq:fQ_ST}, which contains higher-order derivatives of $\xi^\mu$ and therefore signals an Ostrogradski-type instability in the would-be gravitational sector. This is the underlying reason for the unavoidable presence of a ghost (that may not appear in the linear spectrum on strongly coupled backgrounds). Crucially, \emph{this issue is an intrinsic feature of the theory} rather than a mere artifact of the particular backgrounds considered in the perturbative analysis. Indeed, the problem can be understood from the fact that, in the coincident gauge, the theory reduces to a formulation in terms of the metric alone, whose rank-2 gravitational field generically propagates 7 modes (instead of the 2 or 5 associated with a healthy spin-2 field).

In this respect, it is instructive to compare with what occurs in massive gravity and the associated Boulware-Deser ghost. In that context, however, there exists a well-defined procedure to eliminate this intrinsic instability: by appropriately tuning the potential, one arrives at the de Rham-Gabadadze-Tolley formulation \cite{deRham2010}, in which the Boulware-Deser ghost is removed non-perturbatively from the spectrum. Although ghost-like modes can still appear on certain backgrounds (e.g., around cosmological solutions), these are not associated with the original Boulware-Deser ghost, which has been consistently eliminated. In contrast, for $f(\Q)$ theories, no analogous mechanism is currently known that would remove the intrinsic pathological mode. 

We will amuse ourselves with a potential mechanism to render $f(\Q)$ theories viable. We first notice that the action \eqref{eq:fQ_ST} in the coincident gauge $\xi^\alpha=x^\alpha$ can be rewritten in a more suggestive form by using the relations
\begin{align}
q^{\alpha\beta} q^{\mu\nu}\partial_\beta q_{\mu\nu}=\partial^\alpha\log |q|,\quad q^{\alpha\mu} q^{\beta\nu}\partial_\beta q_{\mu\nu}=-\partial_\beta q^{\beta\alpha}.
\end{align}
We can then write the action in the following form:
\begin{align}
    \mathcal{S}_{\text{ST}} =\frac12 \int\dd^4x\,\sqrt{-q}\ &\Big[R(q)+6q^{\mu\nu}\partial_\mu \varphi\partial_\nu \varphi\nonumber\\
    & -2 \big(\partial^\alpha\log |q| +\partial_\beta q^{\beta\alpha}\big)\partial_\alpha\varphi  \nonumber \\
    & - \ee^{-2\varphi} q^{\mu\nu}\partial_\mu \chi \partial_\nu \chi +\mathcal{U}(\varphi, \chi)\Big].
    \label{eq:fQ_ST2}
\end{align}
This form is suggestive because the first term of the Diff-breaking sector $\partial^\alpha\log |q| \, \partial_\alpha\varphi$ respects volume-preserving-Diffs, i.e., it has an additional TDiff symmetry. Then, by using the findings in \cite{Bello-Morales:2024vqk}, where it was shown that TDiff symmetric theories are non-pathological, we can conclude that the pathological character of $f(\Q)$ originates from the coupling $\partial_\beta q^{\beta\alpha} \partial_\alpha\varphi$. Then, it may seem obvious that the theory can be rendered non-pathological by removing that term via e.g. a Lagrange multiplier setting $\partial_\beta q^{\beta\alpha}$ to zero. Of course, the resulting theory is different from the original $f(\Q)$ and its observational or phenomenological viability should be thoroughly analysed, but the main objection for $f(\Q)$ theories would be avoided.

~

\textbf{The coupling to matter}.
Another recurrent claim is that non-minimal couplings or couplings of the connection to matter fields might resolve the pathologies. This probably stems from a misunderstanding of the problem in $f(\Q)$. The Ostrogradski instability that we already discussed is not cured by introducing additional interactions with matter fields.\footnote{
    We have in mind generic couplings. Of course one could always cure the instabilities by e.g. introducing a coupling to a matter field that happens to be a Lagrange multiplier freezing the problematic as in the TDiff-invariant version discussed above.}
In fact, if anything, introducing additional couplings to matter will make the situation even worse in general because it would open new channels of instabilities. In general, we could, at most, consider the theory, including couplings to matter fields, as an effective field theory valid below the scale at which the ghost enters, but this would bring us back to STEGR with some perturbative corrections, analogous to standard GR with higher-order operators.
    
~

\textbf{The role of Hamiltonian analysis}.
A full Hamiltonian analysis is often regarded as the standard tool for counting degrees of freedom and identifying the constraints of a theory. However, in the case at hand, both the number and the nature of the degrees of freedom are already resolved. A Hamiltonian formulation could therefore recast this result in terms of canonical variables and constraints, and confirm the propagation of 7 degrees of freedom, but it would not change the conclusion concerning the instability.

~

\textbf{Particular choices of $f$}. 
It is some times stated that special functional forms of the $f(\Q)$ theory may avoid the ghost issue. A trivial choice that ensures the absence of the ghost is $f=\Q$, but this brings us back to GR. The singular character of the linear function resides in the enhancement of the gauge symmetry to include a second copy of Diffs. Other degenerate functions giving rise to ghost free theories are not currently known. In such degenerate cases one must demonstrate explicitly how the structure of the theory ensures the absence of the problematic degrees of freedom.

~

\textbf{Non-propagating ghost}. It has been repeatedly claimed that the ghost in $f(\Q)$ gravity is non-propagating and therefore harmless~\cite{Hu:2023gui}. The reduced quadratic action shows the opposite: $f(\Q)$ propagates seven degrees of freedom, at least one of them a ghost. The non-propagating interpretation rests on mistaking first time derivatives in selected unreduced equations for a degree-of-freedom count.\footnote{
    The claim originates from the statement below Eq.~(55) of Ref.~\cite{Hu:2023gui}, where $\varphi$ and $A^a_{\ast}$ are said not to propagate because they enter the equations only through first time derivatives. This inference is premature: $\varphi$ enters with second time derivatives in the equations for $\bar A^a_i$, in the notation of Ref.~\cite{Hu:2023gui}. Auxiliary variables must be eliminated and the reduced kinetic matrix analysed before the spectrum can be identified. In this note, we have explicitly shown in the scalar-tensor representation that the full theory can propagate a maximum of 7 dofs. Since the cosmological background saturates this number, we are led to conclude that the full $f(\Q)$ theories generically have 7 propagating degrees of freedom.}

~

\textbf{On curing the ghost}. 
To really rescue $f(\Q)$ theories and make them a viable framework for physical applications, one needs to tackle the ghost problem. It is worth mentioning that there has been recent progress on the stability of theories with ghosts or unbounded Hamiltonians (see e.g. \cite{Deffayet:2021nnt, Deffayet:2023wdg, Deffayet:2025lnj, ErrastiDiez:2024hfq, ErrastiDiez:2025mgu}) so one may be tempted to invoke those results to motivate using $f(\Q)$ theories despite having ghosts. This, however, is not a satisfactory solution, for several reasons. Firstly, the stable systems with ghosts that have been found mostly correspond to either some point-like mechanical systems or very specific field theories so, at most, perhaps some special functional forms of the $f(\Q)$ theories could be stable in certain sense. However, one needs to bear in mind that this theory will be coupled to matter fields and it could have arbitrary perturbations (stemming from matter interactions) so that the stability will be compromised once again. In any case, given the importance of the issue, any mechanism by which the ghost problem is claimed to be avoided must be carefully and thoroughly tackled before phenomenological applications can be taken seriously. An alternative to try to make sense of $f(\Q)$ theories would also be to accept the presence of the ghost and the associated pathologies and introduce some well-founded and consistent prescription to eliminate them. In this respect, one could envision applying the prescription dubbed Purely Virtual Particle (see e.g. \cite{Anselmi:2017yux, Anselmi:2017ygm, Piva:2023bcf}) to the ghost degree of freedom. Of course, this procedure affects not only the ghost, but also the dynamics. As a result, the classical dynamics inferred from the original $f(\Q)$ action cannot be straightforwardly trusted once such a prescription is adopted and, in fact, generically does not correspond to the classical limit of the resulting quantum theory so the phenomenological applications must be reassessed.

\acknowledgments{
The authors would like to thank Laur J\"arv for his encouragement to revisit and complete this work, as well as for his detailed reading and suggestions to improve the manuscript.
J.B.J. was supported by project PID2024-158938NB-I00 (MICIU/AEI/10.13039/501100011033 and ``ERDF -- A way of making Europe") and by project SA097P24 (Junta de Castilla y Le\'on). A.J.C. is supported by the grant PID2022-140831NB-I00 funded by MCIN/AEI/10.13039/501100011033. T.S.K. is supported by the Estonian Research Council grants  TARISTU24-TK10, TARISTU24-TK3, CoE TK202 ``Foundations of the Universe'' and PRG2608 ``Space - Time - Matter''.
The computations have been checked with the xAct package \cite{xAct}.
}

\bibliography{references.bib}

\end{document}